\pdfoutput=1
\documentclass[pdftex,pra,aps,twoside,nofootinbib,showkeys,showpacs,twocolumn,superscriptaddress]{revtex4-2}
\usepackage{setspace}
\usepackage{hyperref}
\usepackage{float}
\usepackage{mciteplus}
\usepackage{indentfirst}
\usepackage{url}
\usepackage[stable]{footmisc}
\usepackage{fancyhdr}
\usepackage{appendix}
\usepackage{epstopdf}
\usepackage[all]{xy}
\usepackage{ifpdf}
\usepackage{color}
\ifpdf
 \usepackage[pdftex]{graphicx}
 \else
\usepackage{graphicx}
 \fi
\usepackage{graphicx}
\usepackage[T1]{fontenc}
\usepackage{textcomp}
\usepackage{newcent}
\usepackage[sc,noBBpl]{mathpazo}
\usepackage{amsmath,amsfonts,amssymb,amsthm, graphics}
\usepackage[mathscr]{eucal}
\usepackage{appendix}
\usepackage{bbm,bm}

\theoremstyle{plain}
\newtheorem{theorem}{Theorem}

\newtheorem{definition}[theorem]{Definition}

\newtheoremstyle{note}{\topsep}{\topsep}{\slshape}{}{\scshape}{}{ }{}
\theoremstyle{note}

\newcommand\tr{\operatorname{Tr}}

\newcommand{\<}{\langle}
\renewcommand{\>}{\rangle}

\newcommand\be{\begin{equation}}
\newcommand\ee{\end{equation}}
\newcommand\bea{\begin{array}}
\newcommand\eea{\end{array}}
\newcommand\ben{\begin{eqnarray}}
\newcommand\een{\end{eqnarray}}

\newcommand\tred{\textcolor{red}}
\newcommand\bei{\begin{itemize}}
\newcommand\eei{\end{itemize}}
\newcommand\bee{\begin{enumerate}}
\newcommand\eee{\end{enumerate}}
\newcommand{\ket}[1]{| #1 \rangle}
\newcommand{\bra}[1]{\langle #1 |}

\newcommand{\mathsym}[1]{{}}
\newcommand{\unicode}[1]{{}}

\def\<{\langle}
\def\>{\rangle}

\def\tred{\textcolor{red}}

\newcommand{\I}{\mathcal I}
\newcommand{\h}{\mathcal H}

\newsavebox{\smlmat}
\savebox{\smlmat}{$\left[\begin{smallmatrix}p&\alpha \\\alpha^{*}&\widetilde{p}\end{smallmatrix}\right]$}

\begin{document}

\title{Roads to objectivity: Quantum Darwinism, Spectrum Broadcast Structures, and Strong quantum Darwinism -- a review}

\author{J. K. Korbicz}
\email{jkorbicz@cft.edu.pl}
\affiliation{Center for Theoretical Physics, Polish Academy of Sciences, 02-668 Warsaw, Poland}
\date{\today}

\begin{abstract}
The problem of objectivity, i.e. how to explain on quantum grounds the objective character of the macroscopic world, is one of the aspects of the celebrated quantum-to-classical transition. Initiated by W. H. Zurek and collaborators, this problem gained some attention recently with several approaches being developed.  The aim of this work is to compare three of them: quantum Darwinism, Spectrum Broadcast Structures, and strong quantum Darwinism. The paper is concentrated  on foundations, providing a synthetic analysis of how the three approaches realize the idea of objectivity and how they are related to each other. As a byproduct of this analysis, a proof of a generalized Spectrum Broadcast Structure theorem is presented. Recent quantum Darwinism experiments are also briefly discussed. 

\end{abstract}

\maketitle
\section{Introduction}

The fact that classical limit of quantum theory should not only reproduce the correct kinematics and dynamics of classical theory, but
on a deeper level also its objective character, has been noted fairly late, at least to the author's knowledge, by W. H. Zurek and collaborators (see e.g. \cite{Ollivier, Zurek_poczatki, Zurek Nature} and the references therein). Quantum measurements inevitably perturb measured system, unless the state of the system is specially tailored for a given measurement, thus precluding in general any objective character of the measured quantity as different observations will yield different results. This is in a stark contrast to classical mechanics, where system's characteristics, e.g. positions or momenta, have an objective, observer-independent character. This  problem of objectivity, unlike the well studied problems of uncertainty and contextuality, arises already at the level of a single observable. 
In recent years several objectivization mechanism have been proposed, in particular
quantum Darwinism \cite{Zurek Nature}, Spectrum Broadcast Structures (SBS) \cite{PRL, PRA}, and strong quantum Darwinism \cite{Le}. The aim of this work is to analyze these ideas from a foundational point of view and study their mutual interrelations in an attempt to bring more clarity to the growing field. 

To set the ground, it is worth to first define what "objectivity" means.  A particularly straightforward, and to some extent operational, definition has been proposed in \cite{Zurek Nature}, namely:

\begin{definition}[Objectivity]\label{def:obj}
A state of the system $S$ exists objectively if many observers can find out the state
of $S$ independently, and without perturbing it.
\end{definition}

It is worth stressing that the notion of objectivity used here concerns the state of the system and not the system itself. Definition \ref{def:obj} does not give a hint on how the objectivization process happens and this has to be added separately. Here comes the idea of redundant information proliferation \cite{Zurek Nature}: Information about the state of the system is multiplied and deposited in many identical records during the dynamical evolution. If it then can be independently accessed without a disturbance, whatever that may mean, it becomes objective. This heuristic idea, central to quantum Darwinism, is actually at the core of all the three approaches. It is in its implementation where they differ. 

If one accepts the redundant information proliferation as a possible explanation of the perceived objectivity, then those multiple records of information must be physically deposited somewhere. The natural candidate is an environment. Every real-life system is immersed in some sort of environment and most of our everyday observations are in fact indirect, mediated by the environment. This is the key difference between the objectivity studies and the standard quantum open systems theory \cite{open systems}. The environment here is promoted to a valuable information carrier and is not a mere source of dissipation and noise. As a result, we consider the following physical setup \cite{Zurek Nature}: The system of interest $S$ (the central system) interacts with an environment $E$, itself composed out of a number of subsystems $E_1,\dots, E_N$. Some fraction of the environment, denoted $fE$ and meaning a collection of $fN$, $0<f<1$ subenvironments, is assumed to be under observation and thus cannot be traced out as it is usually done. This part is where we will be looking for information about the system. The rest, denoted $(1-f)E$, escapes the observation and thus can be traced out, inducing decoherence. This setup is an refinement of the standard framework of open systems and decoherence in that: i) the environment is structured and not treated as a solid block;  ii) parts of the environment are observed as they can carry a valuable information about the system.  

In what follows, I will use the above general setup and discuss the three approaches to objectivity, starting with historically the first one, quantum Darwinism, in Section \ref{sec:qD}. Then I will make an important detour and elaborate on the notion of non-disturbance in Section \ref{Sec:nondisturb} to come back to quantum Darwinism and the role of conditional quantum mutual information in Section \ref{Sec:cmi}. Section \ref{Sec:SBS} discusses Spectrum Broadcast Structures (SBS), including a generalized SBS theorem. Strong quantum Darwinism is presented in Section \ref{Sec:sqD}. In Section \ref{sec:exp}, I briefly discuss recent experiments, aimed at simulating/checking for quantum Darwinism. 

\section{Quantum Darwinism and the meaning of quantum mutual information}\label{sec:qD}
As mentioned in the Introduction, the main idea behind quantum Darwinism is that in the course of the interaction with the environment, some information about the system's state not only "survives" the interaction but also "multiplies" in the environment in the sense that it becomes: i) redundantly encoded in many copies in the environment fragments and ii) locally accessible for a readout without disturbance, thus becoming objective. 
To check if that has taken place,  Zurek and collaborators proposed an entropic condition  \cite{Zurek Nature}, which in its sharp (ideal) form can be stated as follows: 
\begin{definition}[Quantum Darwinism]
There exists an environment fraction size $f_0$ such that for all fractions larger than it, $f\geq f_0$, it holds:
\begin{equation}\label{qD}
\I(\varrho_{S:fE})=\h(\varrho_S)\  \text{\it independently of }  f.
\end{equation}
\end{definition}
Before proceeding with the analysis of the above Definition, it is a good place to introduce the basic notation. Following the usual quantum information practice, quantum states will in general carry indices of their corresponding Hilbert spaces. For example, $\varrho_S$ in a quantum state on the system's Hilbert space $\h_S$; multipartite states will be denoted by $\varrho_{A:B}$, meaning a state on a tensor product $\h_A\otimes \h_B$ and the colon denotes which partition we are interested in.
The function $\h(\varrho)\equiv-\tr(\varrho\log\varrho)$ is the von Neumann entropy.
Finally, $\I(\varrho_{S:fE})$ is the quantum mutual information between the system $S$ and a fraction $fE$ of the environment, defined as \cite{Nielsen Chuang}:
\be\label{qmi}
\I(\varrho_{A:B})\equiv\h(\varrho_A)+\h(\varrho_B)-\h(\varrho_{A:B}),
\ee
where $\varrho_{A,B}\equiv \tr_{B,A}\varrho_{A:B}$ are reductions (partial traces) of the bipartite state $\varrho_{A:B}$ to respective subspaces $A$ and $B$. In what follows, a simplified notation $\I(A:B)$ will be used instead of $\I(\varrho_{A:B})$ whenever this will not lead to a confusion.

Condition (\ref{qD}) is usually motivated as follows: Since quantum mutual information captures total correlations present in the state, for the moment interpreted intuitively as some  sort of information, and enlarging the fraction size past some critical size $f_0$ brings no new correlations, this means that fraction $f_0$ already has the full information about the system, as measured by the entropy of the reduced state $\h(S)$. It is then tacitly assumed that any environment fraction of the size $f_0$ has this property and since there are $R_0=N/(f_0 N)=1/f_0$ such disjoint fractions, the information is redundantly encoded in $R_0$ copies, the number being called information redundancy. 

Of course one cannot expect (\ref{qD}) to hold for arbitrary large fractions as for bigger and bigger fractions more correlations are taken into account, including quantum ones involving large portions of the environment (e.g.  for pure global states $\I(S:E)=2\h(S)$). So in fact the quantification "for all" in (\ref{qD}) is not exactly precise as (\ref{qD}) can possibly be expected to hold in some window $f_0\le f\le f_{max}$ only, but this is not important here. Also in realistic situations one would expects some errors and substitute the ideal condition (\ref{qD}) with a softened version, allowing for some relative error (called deficit) $\delta$: 
\be\label{dqD}
\h(S)\ge\I(S:fE)\ge (1-\delta)\h(S).
\ee 
The most direct way to check for this condition in concrete situations is via so called partial information plots (PiP's), where $\I(S:fE)$ (usually averaged over all environment fractions of a size $f$) is plotted as a function of $f$. If the plot shows a characteristic plateau for some range of $f$, called classicality plateau, then ($\delta$-approximate) objectivity of the pointer states \cite{pointers} is concluded \cite{Zurek_poczatki}. And indeed the above condition has been shown to hold in a variety of fundamental models of open quantum systems, including e.g. Collisional decoherence \cite{Zurek collisional}, Quantum Brownian Motion \cite{Zurek qbm, Paz}, Spin-spin systems \cite{Zurek spins} and more. There have been obtained also some interesting general results e.g. in \cite{Zurek SciRep, Amplification, Piani}. Quantum Darwinism literature is quite numerous and this is just a small sample (for the most recent research see e. g. \cite{recent}). Recently there appeared also experiments aimed at observing (\ref{dqD}) \cite{Ciampini, Chen, Jelezko}, which I discuss separately in Section \ref{sec:exp}.

The above approach has been however constructively criticized first in \cite{PRA} and later in \cite{Le} as not always reflecting what it is supposed to. In what follows,  I will elaborate in more detail and extend the arguments from \cite{PRA}, since the original argumentation was quite brief and with certain omissions making it perhaps not very clear. 

Let us first look at the quantum mutual information alone as it is central to (\ref{qD}). In what sense it quantifies the total amount of correlations and represents "what the fragment $fE$ and $S$ know about each other"?  Of course we have a mathematical fact
that quantum mutual information vanishes if and only if the state is completely uncorrelated, i.e.:
\be
\I(A:B)=0 \text{ iff } \varrho_{A:B}=\varrho_A\otimes\varrho_B,
\ee
but if we agree that ultimately objectivity and conditions to test it should posses an operational meaning, the things are not as straight forward as in classical information theory.  It is a good place to recall the existing operational meanings of quantum mutual information. The first one \cite{Groisman} is that $\I(A:B)$ quantifies the asymptotically minimal amount of local noise needed to erase all correlations in a given state $\varrho_{A:B}$. More precisely, define adding local noise through a randomizing  map, say on the  $B$-side, via:
\be
\sigma \longrightarrow \frac{1}{N}\sum_{i=1}^N p_i ( {\mathbf 1} \otimes U_i) \sigma  (\mathbf 1 \otimes U_i)^{\dagger},
\ee
where $p_i$ are probabilities and $U_i$ a collection of unitary matrices. Then the central and technical result of \cite{Groisman} states that for sufficiently large number of copies $n$ of the state $\varrho_{A:B}$, one can always find the randomization map so that it brings $\varrho_{A:B}^{\otimes n}$ arbitrary close to the $n$ copies of the completely uncorrelated state $\varrho_A\otimes\varrho_B$, and it uses a number of: 
\be\label{logN}
\log N \simeq n\I(A:B)
\ee 
unitaries. There is also a symmetric version of this theorem, using exchange entropy rather then the randomizing ensemble length to quantify the noise, but it is not relevant here as the result is the same. The point is that this operational definition does not really seem to be compatible with what we are aiming at: Instead of extracting information about one system by measuring the other it describes a complete destruction of all correlations by ignorance. 

The second operational definition of $\I(A:B)$ identifies it as, so called, entanglement-assisted classical capacity of a quantum channel \cite{Bennett}. A quantum channel $\Lambda$  is completely positive, trace preserving map, i.e. $\tr\Lambda(\sigma)=\tr\sigma$ and $\sigma\geq 0 \Rightarrow {\bf 1}\otimes \Lambda(\sigma) \geq 0$ for an arbitrary extension by unity $\bf 1$. The operational task here is the following: Two parties share $n$ copies of an entangled state (can be taken pure). One party (say $A$) encodes a classical message of the length $n$ on her subsystems and sends them to $B$ via $n$ independent uses of a noisy quantum channel $\Lambda$. $B$ tries to decode the message using generalized measurements (POVM) on the joint state of the $n$ transmitted subsystems and his half of the pre-established entangled states. Then one can prove tha the capacity of such a scheme is given by: 
\be\label{EA}
C_{ea}(\Lambda)=\max_{\ket {\psi_{A:B}}}\I[(\Lambda\otimes {\mathbf 1})\ket {\psi_{A:B}}],
\ee
where the maximum is taken over bipartite pure states $\ket{\psi_{A:B}}$. Again the result is not easy to show and the interested reader is referred to \cite{Bennett}. This interpretation is much closer in spirit to the "objectivity task", which indeed admits an attractive view as a broadcasting task. Information about the state of the central system $S$ should be faithfully broadcasted in many copies into fragments of the environment by quantum channels $\Lambda^{S\to fE}$, effectively established through the system-environment interaction \cite{Zurek Nature}; symbolically:
\be
\Lambda^{S\to fE} (\varrho_S) \equiv \tr_{E \backslash fE}\varrho_{S:E}=\varrho_{fE},
\ee
where $\tr_{E \backslash fE}$ means tracing over the part of $E$ with the fraction $fE$ excluded, so that the resulting state is a state of the fraction $fE$. Although as it will become clear later, this view has been in general fruitful for the objectivity studies, the entanglement-assisted capacity interpretation (\ref{EA}) does not really suit it well as there is no natural pre-established entanglement here. Indeed, it would be rather a highly artificial situation where the observers would have access not only to parts of the environment but also to some "secret" subsystems, entangled with the central system. 

Thus neither of the operational procedures seem suited for the objectivity setup. This suggest that quantum mutual information may not be the best suited figure of merit here. The view is further reinforced if we look from the perspective of the operational task of interest, i.e. information extraction through local measurements. It is connected to measurement-induced disturbance and so called quantum discord (see e.g. \cite{Zurek discord, Modi}), which I discuss in the next Section.

\section{Non-disturbance, EPR-Bohr debate and objectivity}\label{Sec:nondisturb}
Consider two observers performing generalized measurements $M^A=\{M^A_i\}$ and  $M^B=\{M^B_j\}$ on their respective subsystems. Recall that a generalized measurement is defined as a collection of Hermitian operators $M=\{M_i\}$, each corresponding to a given result $i$, and satisfying $M_i\geq 0$, $\sum_i M_i={\mathbf 1}$. The probability of occurrence of the result $i$ when measuring $M$ is then given by $p(i|M)=\tr(M_i\varrho)$. In case of the two observers, they obtain a joint probability distribution of the results according to:
$p(ij|M^A,M^B)=\tr(M^A_i\otimes M^B_j \varrho_{A:B})$ and can calculate the classical mutual information for this distribution $I(M^A:M^B )\equiv I(p(ij|M^A,M^B))$, describing the observed correlations between the measurement results. 
Recall that classical mutual information for a joint probability distribution $p_{ij}$ is defined as $I(p_{ij})\equiv H(p_i)+H(p_j)-H(p_{ij})$, where $H(p_i)\equiv -\sum_ip_i\log p_i$ is the Shannon entropy and $p_{i,j}$ are the marginals of $p_{ij}$.
It is then a well known fact \cite{Luo, Aolita} that for almost all quantum states, i.e. except for a set of measure zero, the quantity known as two sided discord $D(A:B)$ satisfies:
\be\label{D2}
D(A:B)\equiv \I(A:B) - \max_{M^A, M^B} I(M^A:M^B) > 0,
\ee
i.e. information extractable via local measurements, as quantified by $I(M^A:M^B)$,  is strictly less than the quantum mutual information. Moreover, as proven in \cite{Luo, Modi}, local information extraction typically disturbs the whole state in the sense that:
\ben
&& \exists \{\Pi_i^A\}, \{\Pi_j^B\} \text{ s.t. }  \sum_{i,j} \Pi_i^A\otimes\Pi_j^B \varrho_{A:B} \Pi_i^A\otimes \Pi_j^B = \varrho_{A:B} \nonumber\\
&& \Leftrightarrow D(A:B)=0 \Leftrightarrow \varrho_{A:B}=\sum_{i,j}p_{ij} \Pi_i^A\otimes \Pi_j^B \label{nondisturb}
\een
where $p_{ij}$ is some two-party probability distribution and $\{\Pi_i^A\equiv\ket i \bra i\}, \{\Pi_j^B\equiv\ket j \bra j\}$ are one-dimensional projective  (von Neumann) measurements, i.e. measurements associated with some orthonormal bases $\{\ket i\}$ and $\{ \ket j\}$. A notation remark: $\Pi_i^A$ will always denote a projective measurement, acting on space $A$ with the results indexed by $i$. One-dimensionality, also called sharpness, is crucial here as otherwise the above fact will not hold. Some of the implications in (\ref{nondisturb}) are easy to prove. For example, if the state is of the form $\varrho_{A:B}=\sum_{i,j}p_{ij} \Pi_i^A\otimes \Pi_j^B$, then it satisfies $\sum_{i,j}\Pi_i^A\otimes\Pi_j^B \varrho_{A:B} \Pi_i^A\otimes \Pi_j^B =\sum_{ijkl}p_{kl} \Pi_i^A\Pi_k^A\Pi_i^A\otimes\Pi_j^B \Pi_l^B\Pi_j^B=\varrho_{A:B}$, since $\Pi_i^A\Pi^A_j=\delta_{ij}\Pi_i^A$ (and similarly for $B$). Moreover, for such $\varrho_{A:B}$ quantum mutual information $\I(A:B)$ equals the classical one $I(p_{ij})$, since $\h(\varrho_{A:B})=-\tr(\varrho_{A:B}\log\varrho_{A:B})=-\tr(\sum_{i,j}(p_{ij}\log p_{ij}) \Pi_i^A\otimes \Pi_j^B)=H(p_{ij})$ (the Shannon entropy) and similarly for the reduced states $\varrho_A,\varrho_B$. Hence $D(A:B)=0$. The rest is a bit more difficult to show. As a remark, the magnitude of discord will not play any role here, only if it is non-zero.

Now it becomes clear where the problem with the quantum mutual information lies
in the objectivity setting: Eq. (\ref{D2}) implies that the correlations between measurements on the environment fragments and directly on $S$ (performed by some reference observer) will be typically strictly less than $\I(S:fE)$, even if we optimize the measurements. Moreover (\ref{nondisturb}) says that they will in general disturb the state. Thus the presence of discord prevents operational checking for (\ref{qD}) using local measurements.
Relaxing to one-sided measurements, 
will not help as the results (\ref{D2}), (\ref{nondisturb}), including genericity, still hold for the one-sided discord \cite{Aolita}: 
\be\label{d}
\delta(A|B)\equiv\I(\varrho_{A:B}) - \max_{M^B} \I[({\bf 1}\otimes M^B) \varrho_{A:B}],
\ee
in particular: 
\ben
&& \exists \{\Pi_j^B=\ket{j}\bra j\} \text{ s.t. }  \sum_{j} {\mathbf 1}\otimes\Pi_j^B \varrho_{A:B} {\mathbf 1}\otimes \Pi_j^B = \varrho_{A:B}, \nonumber\\
&&  \Leftrightarrow \varrho_{A:B}=\sum_{j}p_{j} \varrho_j^A\otimes \Pi_j^B  \Leftrightarrow \delta(A|B)=0 \label{nondisturb_one}
\een
As above, some of the implications here are easy to see due to the orthogonality of the von Neumann measurements, others being more involved.


As discord is related to measurement-induced disturbance, it is worth to elaborate more on the notion of disturbance, somewhat repeating the arguments of \cite{PRA}, as it will be useful in later considerations. The notion of disturbance discussed so far seems more strict than the Definition ~\ref{def:obj} requires, since it concerns the whole state $\varrho_{S:fE}$. 
This is an important point where objectivity meets the very foundations of quantum mechanics -- the famous EPR-Bohr debate on the completeness of quantum mechanics \cite{EPR, Bohr}.  The notion of disturbance plays a pivotal role in Bohr's rebuttal of the EPR conclusion that quantum mechanics in incomplete. Requiring that the state of the system alone is not disturbed by measurements on the environment is always trivially satisfied, since quantum mechanics obeys the non-signaling principle:
\be\label{non-sig}
\tr_E\varrho_{S:E}=\tr_E [({\mathbf 1} \otimes \Lambda) \varrho_{S:E}],
\ee
for any  quantum operation (trace-preserving completely positive map)  $\Lambda$, including of course measurements. The most straightforward way to see it is to decompose $\varrho_{S:E}$ in arbitrary bases of $S$ and $E$. Then the RHS equals
 $\sum_{ijkl}\varrho_{ijkl}\tr_E [({\mathbf 1} \otimes \Lambda) \ket i\bra j\otimes \ket k\bra l]=\sum_{ijkl}\varrho_{ijkl}\tr_E [\ket i\bra j\otimes \Lambda(\ket k\bra l)]=\sum_{ijkl}\varrho_{ijkl}\ket i\bra j \langle l | k\rangle=\sum_{ijk}\varrho_{ijkk}\ket i\bra j =\tr_E\varrho_{S:E}$ where the trace preserving property of $\Lambda$ was used.
 Property (\ref{non-sig}) may be regarded as a modern  reading of the EPR notion of non-disturbance (a "mechanical disturbance" in Bohr's words), or at least a sufficient condition for it \cite{Wiseman}. But Bohr found it too narrow and postulated that a measuring procedure disturbs the system if only there is an "influence on the very conditions which define the possible types of predictions regarding the future behavior of the system" \cite{Bohr}. While this statement, much like the whole of Bohr's reply, has been notoriously difficult to follow, Wiseman \cite{Wiseman} has formalized it recently in modern operational terms. He has identified that what "defines the possible types of predictions" is the whole joint quantum state, not just $\varrho_S$, and a measurement has no influence on the "the possible types of predictions regarding the future behavior of the system" only if it preserves the joint state. The idea is that although no measurement on $E$ will change the reduced state of $S$ by the property (\ref{non-sig}), it will in general alter results of possible future measurements on the same system $E$ and thus will influence what  can be predicted about $S$ by measuring $E$. Admittedly it is a very subtle form of disturbance, where mere influence on possible predictions about the state
amounts to disturbing it. But if one accepts Bohr's arguments, one has to accept it. We thus arrive at the following definition of non-disturbance \cite{Wiseman}:

\begin{definition}[Bohr non-disturbance]\label{Bohr}
A measurement $\{\Pi^E_j\}$ on the subsystem $E$ is Bohr non-disturbing on the subsystem $S$ iff:
\be\label{eq:Bohr}
\sum_j \left({\mathbf 1}\otimes \Pi^E_j\right)\varrho_{S:E}\left({\mathbf 1}\otimes\Pi^E_j\right)=\varrho_{S:E}.
\ee
\end{definition}
\noindent Clearly, this is a much stronger condition than the non-signaling (\ref{non-sig}). Note that if the measurement $\{\Pi^E_j\}$ consists of one-dimensional projectors, then Bohr non-disturbance (\ref{eq:Bohr}) is equivalent by  (\ref{nondisturb_one}) to non-discordance, which constitutes  the main result of \cite{Wiseman}.

\section{The role of conditional quantum mutual information in quantum Darwinism}\label{Sec:cmi}
So far I have analyzed quantum mutual information and the notion of disturbance without any reference to condition (\ref{qD}). But the right hand side of (\ref{qD}) puts strong mathematical constraints on $\I$ and it can still happen that despite the above objections, these constraints somehow force the discord to vanish. This is to some extent so. It will be helpful to first look at  (\ref{qD}) from the classical information theory point of view, substituting  density matrices with random variables $X$, $Y$, and entropic functions $\h$ and $\I$ with their classical counterparts $H$ and $I$. Let us consider three random variables $X,Y, Y'$ with the corresponding joint probability distribution $p(x,y,y')$.
Then the "redundancy" part of (\ref{qD}), demanding that including more degrees of freedom brings no new information about the system, can be in the simplest form written as: 
\be\label{cqD}
I(X:YY')=I(X:Y),
\ee
where only one new environment, $Y'$, is added. The quantity $I(X:YY')$ is the mutual information between $X$ and the joint variable $YY'$, defined as $I(X\colon YY')\equiv H(X)+H(YY')-H(XYY')$.
As a remark, demanding $I(X:Y)=H(X)$ is of course way to strong classically as it immediately implies that the conditional entropy vanishes $H(X|Y)=0$ or equivalently that the random variables $X$ and $Y$ are deterministically related and the whole problem trivializes.  From the chain rule for the mutual information it follows that  \cite{CoverThomas}:
\be\label{chain}
I(X:YY')=I(X:Y)+I(X:Y'|Y),
\ee
where $I(X \colon Y'|Y)\equiv H(p(x|y))+H(p(y'|y))-H(p(x,y'|y))$ is the conditional mutual information, conditioned on $Y$.
Thus (\ref{cqD}) is simply equivalent to:
\be\label{cond}
I(X:Y'|Y)=0,
\ee
which in turn means nothing else but the conditional independence of $X$ and $Y'$ given $Y$ \cite{CoverThomas}:
\be\label{indep}
p(x,y'|y)=p(x|y)p(y'|y).
\ee
This can be seen by noting that $I(X \colon Y'|Y)=\sum_{x,y,y'}p(x,y,y')\log[p(x,y|y')/p(x|y')p(y|y')]$ and from the convexity of the logarithm, the sum vanishes iff each term vanishes.
Thus in the most direct sense  $Y'$ adds no new information about $X$ other than $Y$. As a consequence, the full probability distribution can be recovered  from conditioning on $Y$: 
\be\label{indep1}
p(x,y,y')=p(y)p(x|y)p(y'|y).
\ee

How does this look like in the quantum case? Let us again assume the simplest situation of adding one subsystem:
\be\label{SEE}
\I(S:EE')=\I(S:E).
\ee
The chain rule (\ref{chain}) extends also to the quantum mutual information, i.e. \cite{Nielsen Chuang}:
\be
\I(S:EE')=\I(S:E)+\I(S:E'|E)
\ee
where the quantum conditional mutual information is defined through von Neumann entropies in a formally the same was as its classical counterpart:
\be
\I(S:E'|E)\equiv \h(SE)+\h(EE')-\h(E)-\h(SEE').
\ee
Thus (\ref{SEE}) can be rephrased similarly to (\ref{cond}) as:
\be\label{qcond}
\I(S:E'|E)=0.
\ee
However, now this condition is much more difficult to study than the classical counterpart (the mere fact that $\I(S:E'|E)\ge 0$ is already a highly non-trivial result, known as strong subadditivity of the von Neumann entropy \cite{Lieb}). Fortunately, the  structure of states satisfying (\ref{qcond}) is known and is given by the following highly non-trivial result \cite{Petz,Hayden}:
\begin{theorem}[Quantum Markov states]\label{Petz:thm}
State $\varrho_{S:EE'}$ satisfies (\ref{qcond}) iff the Hilbert space $\h_E$ can be orthogonally decomposed into "left" and "right" spaces $\h_E=\bigoplus_j \h_j^L \otimes \h_j^R$ s. t.:
\be\label{qindep}
\varrho_{S:EE'}=\bigoplus_j p_j \varrho_{S:E_j^L}\otimes\varrho_{E_j^R:E'}
\ee
where $p_j$ are probabilities and $\varrho_{S:E_j^L}$, $\varrho_{E_j^R:E'}$ are density matrices from $\h_S\otimes \h_j^L$ and $\h_{E'}\otimes \h_j^R$ respectively. 
\end{theorem} 
Eq. (\ref{qindep}) can be seen as a counterpart of (\ref{indep1}). A somewhat distant analogue of (\ref{indep}) is the product form, in the $S:E'$ split, of the projections of $\varrho_{S:EE'}$ onto the each of the $j$-th 
subspace:
\be\label{proj}
\varrho^j_{S:EE'}\equiv\frac{1}{p_j}{\mathbf 1}\otimes \Pi^E_j\varrho_{S:EE'}{\mathbf 1}\otimes\Pi^E_j=\varrho_{S:E_j^L}\otimes\varrho_{E_j^R:E'}.
\ee
where $\Pi^E_j$ are orthogonal projectors, projecting on the subspace $j$ of the sum $\bigoplus_j \h_j^L \otimes \h_j^R$.
Although the measurement $\{\Pi^E_j\}$  is trivially Bohr non-disturbing by the construction (cf. (\ref{qindep},\ref{proj})):
\be\label{Bnondist}
\sum_j {\mathbf 1}\otimes \Pi^E_j\varrho_{S:EE'}{\mathbf 1}\otimes\Pi^E_j=\varrho_{S:EE'},
\ee
it does not reveal the state of the central system. It reveals the sector $j$ in the orthogonal sum (\ref{qindep}), and hence the states $\varrho_{S:E_j^L}$ and $\varrho_{E_j^R:E'}$.
But in general these are not the sates of $S$ (or $E'$) alone since they carry "pieces" of $E$. 
Moreover, since there are no a priori restrictions on the correlations in $\varrho_{S:E_j^L}$ and $\varrho_{E_j^R:E'}$ (they can be e.g. entangled), any attempt to measure the subsystems $E_j^L$ and $E_j^R$ will
generically disturb the central system $S$ and the other environment $E'$ respectively. This is an especially serious consequence for the central system - generically information about its state cannot be extracted without disturbance although there exists a 
Bohr non-disturbing measurement (\ref{Bnondist}).
Obviously, the sistuation is symmetric if we exchange $E$ and $E'$ starting from (\ref{SEE}). 
Thus, although the condition (\ref{qD}) imposes strong constraints on the form of the state $\varrho_{S:fE}$, it does not in general guarantee the possibility of non-disturbant, local information extraction. For that, additional conditions must be added to (\ref{qD}), leading to what is known as strong quantum Darwinism \cite{Le}. It will be discussed in Section \ref{Sec:sqD}.

Summarizing the consideration of the last three Sections, there are some problems with the use of quantum mutual information when checking for objectivity as it: i)  does not seem to have  the right operational interpretation for objectivity; ii) typically contains more correlations than can be extracted locally and without disturbance; iii) the condition (\ref{qD}) is in general too weak to change that. Of course the authors of the quantum Darwinism idea has been well aware that quantum mutual information contains in general more correlations than locally accessible as they are also co-inventors of discord \cite{Zurek discord}. The role of discord in quantum Darwinism is analyzed in \cite{Zurek SciRep}, showing certain complementarity between the locally accessible information (bounded by the Holevo 
quantity $\chi(E|S)$ \cite{Holevo}, which will be introduced later) and the discord in the LHS of (\ref{qD}). Indeed, from definition (\ref{d}) one immediately obtains a decomposition \cite{Zurek SciRep}:
\be\label{qmi decomp}
\I (S:E)=\chi(E|S)+\delta(E|S).
\ee
However, as already mentioned the problem is the very appearance of the discord in the LHS of (\ref{qD}), making (\ref{qD}) effectively non-operational, unless we use global measurements. 
For even if (\ref{qD}) is satisfied, no local and non-disturbing measurements can check for it, unless the discord vanishes, cf. (\ref{D2}, \ref{d}).  This presence of discord has been indeed confirmed in models in \cite{Pleasance,Le2018}. Thus there appears a discrepancy between the operational character of Definition \ref{def:obj} and the condition (\ref{qD}). One may of course ask if Definition \ref{def:obj} should be understood operationally at all. A strong believe that it  should brings us to the next approach.

\section{Spectrum Broadcast Structures (SBS)}\label{Sec:SBS}

In \cite{PRL, PRA} another approach to the objectivity problem was taken. Instead of using somewhat classically motivated conditions for redundant information proliferation, the authors formalized Def. \ref{def:obj} and surprisingly were able to derive a unique state structure compatible with it. I will repat the process below, generalizing it compared to the original derivation \cite{PRA}

Let us look at the phrase 'finding out of the state of $S$ by many independent observers' of Def. \ref{def:obj}.  First of all, the 'state of the system' is understood as one of the eigenstates $\ket i$ of the reduced density matrix of the system:
\be\label{rS} 
\varrho_S=\tr_E\varrho_{S:E}=\sum_ip_i\ket i\bra i. 
\ee 
Thus what we are trying to make objective is $\ket i$, or more precisely its index $i$, since when talking about objectivity of a certain property, i.e. position or momentum, we should first of all know which property we are talking about.
Then 'finding out [...] by many independent observers' is formalized as
observers performing uncorrelated von Neumann measurements, $\{\Pi_{i_k}^k\}$, on their respective subsystems $E_k$. Note that the number of observers need not be specified at all and is in a sense a 'free parameter'. Common sense says that at least two as for one observer everything is objective by definition.  
A notation remark: The subscript $k$ enumerates here the environment subsystems so that $\Pi_{i_k}^k$ is a projector acting in the space of $E_k$ and corresponding to the result $i_k$.
Von Neumann measurements are chosen over generalized ones  as the most informative (due to their orthogonality property).  

It is somewhat of an open question in objectivity studies if a direct observation of the central system $S$
should be allowed or not. In the SBS derivation it is allowed \cite{PRA} and is an important part of the scheme as it acts as a reference measurement, making sure that the indirect ('environmental') observations correspond to the actual state of the system. This closes a certain "intersubjectivity loop" where all the indirect observers agree but their observations are
not referenced to the central system. Indeed such situations may appear in models \cite{Mironowicz_PRA}. An important remark is that the direct measurement does not have to be performed by an agent, it can be a result of a decoherence process. On a technical side, without the direct observation not much can be derived from Def. \ref{def:obj}. We thus assume it. A technical assumption
now comes: The direct observation is maximally resolving, i.e. only 1D measurements $\Pi_i^S=\ket i \bra i$ are allowed on $S$. This echoes the
the vanishing discord conditions (\ref{nondisturb}, \ref{nondisturb_one}). In the language of decoherence theory, the 1D condition corresponds to a complete decoherence,
i.e. no decoherence free subspaces (for a relaxation of this condition see \cite{Mironowicz_PRA}). At the same time, the environmental measurements $\{\Pi_{i_k}^k\}$
can have an arbitrary rank. This reflects the fact that in realistic situations, the observed parts of the environment are large compared to the central system
and many of its degrees of freedom encode the same information about the system. A final assumption on the measurement structure comes from the agreement condition, implicit in Def. \ref{def:obj} : All the observers see the same  state and this is the state that the direct observation reveals, i.e. only events with $i_1=\dots = i_k=\dots =i$ have non-zero probabilities. This agreement condition is natural and needs no special explanation (see \cite{PRA} for a more formal motivation). We thus arrive at the following, completely product,  measurement structure:
\be\label{pomiar}
\ket i \bra i\otimes \Pi_{i}^{1}\otimes\cdots\otimes \Pi_{i}^{k}\otimes\cdots \equiv \ket i \bra i\otimes{\pmb \Pi}_i,
\ee
where ${\pmb \Pi}_i\equiv \Pi_{i}^{1}\otimes\cdots\otimes \Pi_{i}^{k}\otimes\cdots$ denotes a collective measurement operator on the environment, introduced to keep the notation below a bit simpler.

Now comes the crucial moment -- the use of the non-disturbance condition. As already explained in Section \ref{Sec:nondisturb}, if we accept Bohr's 1935 defense of the completeness of quantum mechanics \cite{Bohr} and its formalization in modern terms of non-demolition measurement by Wiseman \cite{Wiseman}, 
then we are bound to use Definition \ref{Bohr} as the definition of non-disturbance. It forces the following condition:
\be\label{central}
\sum_i \left(\ket i \bra i\otimes{\pmb \Pi}_i\right) \varrho_{S:E} \left(\ket i \bra i\otimes{\pmb \Pi}_i\right) =\varrho_{S:E}.
\ee
Note that the measurement $\ket i \bra i$ is non-disturbing on $S$ alone by its construction -- these are by definition the eigenprojectors of $\varrho_S$ (\ref{rS}).
Let us solve the condition (\ref{central}), generalizing the original derivation of \cite{PRA}. Let $\ket{\pmb\phi_\alpha}$ be some basis in the space of all of the environment $E$. An arbitrary state
$\varrho_{S:E}$ can always be decomposed as $\varrho_{S:E}=\sum_{i,j,\alpha,\beta}\varrho_{ij\alpha\beta}\ket i\bra j\otimes\ket{\pmb\phi_\alpha}\bra{\pmb\phi_\beta}$. We then have:
\begin{eqnarray}
&&\sum_i \left(\ket i \bra i\otimes{\pmb \Pi}_i\right) \varrho_{S:E} \left(\ket i \bra i\otimes{\pmb \Pi}_i\right) =\nonumber\\
&&\sum_{i,\alpha,\beta}\varrho_{ii\alpha\beta}\ket i\bra i\otimes{\pmb \Pi}_i \ket{\pmb\phi_\alpha}\bra{\pmb\phi_\beta}{\pmb \Pi}_i 
\end{eqnarray}
Let us now define the following operators on $E$, numbered by $i$ (cf. (\ref{rS})):
\be\label{hatrhoi}
\hat{\pmb\varrho}_i\equiv\sum_{\alpha,\beta}\varrho_{ii\alpha\beta}{\pmb \Pi}_i \ket{\pmb\phi_\alpha}\bra{\pmb\phi_\beta}{\pmb \Pi}_i 
\ee
and their normalized (unit trace) versions ${\pmb\varrho}_i\equiv \frac{1}{\tr\hat{\pmb\varrho}_i}\hat{\pmb\varrho}_i$, $\tr\hat{\pmb\varrho}_i=
\sum_{\alpha,\beta}\varrho_{ii\alpha\beta}\bra{\pmb\phi_\beta}{\pmb \Pi}_i \ket{\pmb\phi_\alpha}$. Then (\ref{central}) implies:
\be
\varrho_{S:E}=\sum_i (\tr\hat{\pmb\varrho}_i) \ket i \bra i \otimes {\pmb\varrho}_i.
\ee
Taking the partial trace with respect to $E$ and using orthogonality of the system states $\ket i$, we may identify $\tr\hat{{\pmb\varrho}_i}$ as the probabilities $p_i$ from (\ref{rS}), i.e. the probabilities of occurrence of states $\ket i$. Taking next the partial trace w.r.t. $S$ and observing that $p_i\geq 0$,
we may easily convince ourselves that for any state $\ket{\pmb \psi}$ from $\h_E$ it holds $\bra{\pmb \psi}{\pmb\varrho}_i \ket{\pmb \psi}\geq 0$, so that ${\pmb\varrho}_i\geq 0$ and are thus legitimate quantum states for every $i$.
Thus, the states satisfying (\ref{central}) must be of the form \cite{PRA}:
\be\label{presbs}
\varrho_{S:E}=\sum_i p_i \ket i \bra i \otimes {\pmb \varrho_i}.
\ee
It is important to note that 
we kept a single, fixed measurement (\ref{pomiar}) and the part of this measurement on the system is 1D; 
for example if $\ket i$ are position eigenstates, measurements $\{\Pi^{k}_i\}$ are some kind of position measurements too. 
But much more can be said about the residual states ${\pmb \varrho_i}$. From the definition (\ref{hatrhoi}) and the orthogonality condition ${\pmb \Pi_i}{\pmb \Pi_j}=\delta_{ij}{\pmb \Pi_i}$, it follows that:
\begin{eqnarray}
&&{\pmb \Pi_i}{\pmb \varrho_i}{\pmb \Pi_i}={\pmb \varrho_i},\label{preserv}\\
&&{\pmb \varrho_i}{\pmb \varrho_j}=0 \ \textrm{for any } i\ne j.\label{pre_perp}
\end{eqnarray}
Although condition (\ref{pre_perp}) follows from (\ref{preserv}), it is convenient to have  both of them written explicitly. 
In particular, (\ref{pre_perp}) means that states ${\pmb \varrho_i}$, ${\pmb \varrho_j}$ have orthogonal supports, i.e. every eigenvector of ${\pmb \varrho_i}$ is orthogonal to every eigenvector of  ${\pmb \varrho_j}$, 
denoted by ${\pmb \varrho_i}\perp{\pmb \varrho_j}$. This has interesting consequences from the quantum information point of view: States ${\pmb \varrho_i}$ are, so called, one-shot distinguishable \cite{Nielsen Chuang}, i.e. 
it is possible to construct a measurement which upon a single application will tell with absolute certainty with which state ${\pmb \varrho_i}$ we are dealing. Obviously, the operators ${\pmb \Pi_i}$ define such a measurement
here. Thus distinguishability property allows to unambiguously recover the index $i$, and hence the state of $S$,  from the environment.
The fully product nature (\ref{pomiar}) of ${\pmb \Pi_i}$, implies an even stronger set of conditions -- the preservation (\ref{preserv}) and one-shot distinguishability (\ref{pre_perp}) hold for all the reductions of ${\pmb \varrho_i}$ to subenvironments.
Namely, consider discarding all the environments apart from some chosen subset of $n$, say $E_{k_1},\dots,E_{k_n}$. Define a reduction of ${\pmb \varrho_i}$  to the space $\h_{E_{k_1}}\otimes \cdots \otimes \h_{E_{k_n}}$ through the partial trace:
\be\label{rhok1kn}
{\pmb \varrho}_i^{k_1\dots k_n}\equiv \tr_{E_1\dots\hat E_{k_1}\dots \hat E_{k_n}\dots E_N}{\pmb \varrho}_i,
\ee
i.e. we trace over all the environments apart from  $E_{k_1},\dots,E_{k_n}$ (indicated by the hats). It then follows immediately from (\ref{pomiar}), (\ref{hatrhoi}), and (\ref{rhok1kn}) that any reduction ${\pmb \varrho}_i^{k_1\dots k_n}$: i) is preserved by the measurement $\Pi_{i}^{k_1}\otimes\cdots\otimes \Pi_{i}^{k_n}$ and ii) has orthogonal supports for different indices $i$ \cite{PRA}. Since the subset  $E_{k_1},\dots,E_{k_n}$ was arbitrary, the above properties hold for any subset. More formally, for any $n=1,\dots, N$ and any subset $k_1,\dots, k_n$ it holds:
\begin{eqnarray}
&&\Pi_{i}^{k_1}\otimes\cdots\otimes \Pi_{i}^{k_n}{\pmb \varrho}_i^{k_1\dots k_n}\Pi_{i}^{k_1}\otimes\cdots\otimes \Pi_{i}^{k_n}={\pmb \varrho}_{i}^{k_1 \dots k_n},\label{preserv2}\\
&&{\pmb \varrho}_i^{k_1\dots k_n}{\pmb \varrho}_{j}^{k_1 \dots k_n}=0 \ \text{for any } i\ne j\label{perp}
\end{eqnarray}
These are stronger, "fine-grained", versions of (\ref{preserv},\ref{pre_perp}). To prove the above, consider the environment basis $\ket{\pmb\phi_\alpha}$ in (\ref{hatrhoi}) to be a product one (always possible from the definition of tensor product),
$\ket{\pmb\phi_\alpha}=\ket{\alpha_1\cdots\alpha_N}$; $\alpha$ is a multiindex $\alpha_1\cdots\alpha_N$. Then from (\ref{hatrhoi},\ref{rhok1kn}), 
${\pmb \varrho}_i^{k_1\dots k_n}=\frac{1}{p_i}\sum_{\alpha,\beta}\varrho_{ii\alpha\beta}\left(\Pi_{l\ne\{k_1,\dots,k_n\}}\langle\alpha_l|\Pi_i^l|\beta_l\rangle\right) \Pi_{i}^{k_1}\otimes\cdots\otimes \Pi_{i}^{k_n}\ket{\alpha_{k_1}\cdots\alpha_{k_n}}\bra{\beta_{k_1}\cdots\beta_{k_n}}\Pi_{i}^{k_1}\otimes\cdots\otimes \Pi_{i}^{k_n}$. From this and the orthogonal character of the von Neumann measurements, the equations (\ref{preserv2}, \ref{perp}) follow easily. 
The immediate
consequence of (\ref{perp}) is that all the reductions  ${\pmb \varrho}_i^{k_1\dots k_n}$ are one-shot perfectly distinguishable for different $i$. The distinguishing 
measurements are of course $\{\Pi_i^{k_1}\otimes\cdots\otimes \Pi_i^{k_n}\}$, which now project on the orthogonal supports 
of ${\pmb \varrho}_i^{k_1\dots k_n}$ for different $i$. Summarizing the above analysis, we obtain the implication: 
\be\label{impl}
\text{Definition \ref{def:obj} $\Longrightarrow$ state (\ref{presbs}) with property (\ref{perp})}.
\ee

Let us now analyze the implication in the opposite direction. Let us assume (\ref{presbs}) and (\ref{perp}) to hold. I will show that such states fulfill Definition 1.
Fix a subenvironment $E_k$. Then by  (\ref{perp}) with $n=1$, the states ${\pmb \varrho}_i^k$, i.e. the reductions of ${\pmb \varrho}_i$ to $E_k$, will have orthogonal supports for different $i$. Define measurements $\{\Pi_i^{k}\}$ as a collection of of orthogonal projectors on these supports, $\Pi_i^{k}\Pi_j^{k}=\delta_{ij}\Pi_i^{k}$.  Repeat the procedure for every subenvironment $E_1, \dots, E_N$. This defines a measurement on every subsystem $E_k$ and these will be the measurements used by the observers to learn the state of $S$. 
Let us first check if there is an agreement among the observers, i.e. if all will observe the same outcome. As before, let us choose an arbitrary subset of environments, say labeled by $\{k_1\dots k_n\}$, on which the measurements will be performed. On the rest of $E$ no action is performed.  Consider the probability of observing result $i$ on the central system $S$, result $i_1$ on $E_{k_1}$, $i_2$ on $E_{k_2}$, and so on:
\begin{eqnarray}
&&p(i,i_1,\dots,i_n)=\tr\left(\varrho_{S:E}\ket i\bra i \otimes \Pi_{i_1}^{k_1}\otimes\cdots\otimes\Pi_{i_n}^{k_n}\otimes{\pmb 1}\right)\nonumber\\
&& =p_i\tr\left(\varrho_i^{k_1\dots k_n}\Pi_{i_1}^{k_1}\otimes\cdots\otimes\Pi_{i_n}^{k_n}\right),
\end{eqnarray}
where (\ref{presbs}) and (\ref{rhok1kn}) were used and ${\pmb 1}$ is the identity operator on those subenvironments which are not measured. Using the cyclic property of the trace and the projective character of the measurement,
we can "isolate" one of the measurements, e.g. the first one, and write $\tr\left(\varrho_i^{k_1\dots k_n}\Pi_{i_1}^{k_1}\otimes\cdots\otimes\Pi_{i_n}^{k_n}\right)=\tr\left[\left(\Pi_{i_1}^{k_1}\otimes{\pmb 1}\varrho_i^{k_1\dots k_n}\Pi_{i_1}^{k_1}\otimes{\pmb 1}\right){\pmb 1}_{k_1}\otimes\Pi_{i_2}^{k_2}\cdots\otimes\Pi_{i_n}^{k_n}\right]$. Noting that the operator in round brackets and the projection after it are both positive semidefinite, we can use a well known trace inequality 
\be
0\leq \tr(AB)\leq \tr A \tr B, 
\ee
valid for $A,B\geq 0$. It then gives
$0\leq \tr\left[\left(\Pi_{i_1}^{k_1}\otimes{\pmb 1}\varrho_i^{k_1\dots k_n}\Pi_{i_1}^{k_1}\otimes{\pmb 1}\right){\pmb 1}\otimes\Pi_{i_2}^{k_2}\cdots\otimes\Pi_{i_n}^{k_n}\right]\leq \tr\left(\Pi_{i_1}^{k_1}\otimes{\pmb 1}\varrho_i^{k_1\dots k_n}\Pi_{i_1}^{k_1}\otimes{\pmb 1}\right) \tr\left({\pmb 1}\otimes\Pi_{i_2}^{k_2}\cdots\otimes\Pi_{i_n}^{k_n}\right)=\tr(\varrho_i^{k_1}\Pi_{i_1}^{k_1}) D$, where $D$ is some strictly positive constant, determined by the dimensionalities of the projectors. But by the construction $\tr(\varrho_i^{k_1}\Pi_{i_1}^{k_1})$ is non-zero only for $i_1=i$. Applying sequentially the above procedure to the remaining measurements, we find that:
\be
p(i,i_1,\dots,i_n)\sim \delta_{ii_1}\cdots\delta_{ii_n},
\ee
that is $p(i,i_1,\dots,i_n)\ne 0$ only for $i=i_1=\cdots=i_n$,  i.e. if observers measure, they all obtain the same result $i$. If the central observer is not involved, the reasoning is similar. Then $p(i_1,\dots,i_n)=\tr\left(\varrho_{S:E}\,{\pmb 1}_S\otimes\Pi_{i_1}^{k_1}\otimes\cdots\otimes\Pi_{i_n}^{k_n}\otimes{\pmb 1}\right)=\sum_i p_i\tr\left(\varrho_i^{k_1\dots k_n}\Pi_{i_1}^{k_1}\otimes\cdots\otimes\Pi_{i_n}^{k_n}\right)$, which by the same iterative procedure can be non-zero only for $i_1=\cdots=i_n$.
This establishes the agreement -- a necessary condition for objectivity. 

To prove non-disturbance, I first show that (\ref{preserv2}) holds. Recall that the measurements $\{\Pi_i^{k}\}$ are defined as the projectors on the supports of the 1-party reductions $\varrho_i^k$. As such, their tensor products leave the states $\varrho_i^{k_1\dots k_n}$ invariant. The easiest way to see it is to first consider pure states, forgetting the index $i$ for the moment. Let $\ket{\psi}=\sum_{l_1 \dots l_n} \psi_{l_1\dots l_n} \ket{l_1 \dots l_n}$ be an $n$-party pure state. Then its 1-party reductions are $\varrho^k=\sum_{l_1 \dots l_n, l'_k} \psi_{l_1\dots l_k \dots l_n} \psi^*_{l_1\dots l'_k \dots l_n} \ket{l_k}\bra{l'_k}$ and are supported on $\text{span}\{\ket{l_k}\}$. The projectors on these supports are given by $\Pi^k=\sum_{l_k}\ket{l_k}\bra{l_k}$, where the sum includes only those basis vectors $\ket{l_k}$ which appear in $\ket{\psi}$. It is now straightforward to see that :
\begin{eqnarray}
\Pi^1\otimes \dots \otimes \Pi^n \ket{\psi} &=& \sum_{l_1 \dots l_n} \psi_{l_1\dots l_n} \Pi^1\otimes \dots \otimes \Pi^n \ket{l_1 \dots l_n}\nonumber \\
&=&\ket{\psi}. 
\end{eqnarray}
For mixed states, consider any decomposition $\varrho=\sum_r \lambda_r \ket{\psi_r} \bra{\psi_r}$.  Now each 1-party reduction is supported on a span which runs also across the ensemble index $r$: $\text{supp}\varrho^k=\text{span}_{r,l^r_k}\{\ket{l^r_k}\}$. There will be in general some linearly dependent vectors for different $r$'s but this does not matter as one can always find a orthonormal set among them and define $\Pi^k$ using it. Then by construction $\Pi^1\otimes \dots \otimes \Pi^n \varrho \Pi^1\otimes \dots \otimes \Pi^n =\varrho$. Repeating this argument in each of the orthogonal, by the assumption (\ref{perp}), sectors $i$, we arrive at (\ref{preserv2}). Thus the  condition  (\ref{perp}) in fact implies (\ref{preserv2}).

The last step is to show that  (\ref{preserv2}) implies Bohr non-disturbance. Consider as before  an arbitrary subset of environments, labeled by $\{k_1\dots k_n\}$, on which  the measurements will be applied. The rest of the environment is left untouched. The average post-measurement state then satisfies:
\begin{eqnarray}
&&\sum_i \left(\Pi_{i}^{k_1}\otimes\cdots\otimes \Pi_{i}^{k_n}\otimes {\pmb 1}\right)\varrho_{S:E}\left(\Pi_{i}^{k_1}\otimes\cdots\otimes \Pi_{i}^{k_n}\otimes{\pmb 1}\right)\nonumber\\
&& =\sum_{i,j} p_{j} \ket{j}\bra{j} \otimes \nonumber\\
&& \otimes \left(\Pi_{i}^{k_1}\otimes\cdots\otimes \Pi_{i}^{k_n}\otimes {\pmb 1}\right){\pmb \varrho_{j}}\left(\Pi_{i}^{k_1}\otimes\cdots\otimes \Pi_{i}^{k_n}\otimes{\pmb 1}\right)\\
&& =\sum_{i} p_{i} \ket{i}\bra{i} \otimes \nonumber\\
&& \otimes \left(\Pi_{i}^{k_1}\otimes\cdots\otimes \Pi_{i}^{k_n}\otimes {\pmb 1}\right){\pmb \varrho_{i}}\left(\Pi_{i}^{k_1}\otimes\cdots\otimes \Pi_{i}^{k_n}\otimes{\pmb 1}\right)\\
&& = \sum_{i} p_{i} \ket{i}\bra{i} \otimes {\pmb \varrho_{i}}=\varrho_{S:E},
\end{eqnarray}
where in the first step (\ref{presbs}) was used. Next, by construction $\Pi_i^k$ project on the subspace orthogonal to the support of  ${\pmb \varrho_j}$ for  $i \ne j$ implying that the only non-zero terms in the double sum are those with $i=j$. In the final step (\ref{preserv2}) was used again, trivially extended by the identity to the whole ${\pmb \varrho_{i}}$.  This concludes the proof of Bohr non-disturbance and hence the reverse implication to (\ref{impl}). The above proof can be repeated without a change when  the direct observer measures as well in the basis $\ket i $.

Summarizing, thanks to the structure (\ref{presbs},\ref{perp}), each of the observers (or groups of them) will obtain the same result $i$, identical to what the direct observer will measure, thus unambiguously identifying the state of $S$.  
Moreover, the whole state $\varrho_{S:E}$ will be unchanged by such measurements once the results are forgotten (irrespectively if all of the measurements are performed or only some), reproducing (\ref{central}) as a consequence.
Thus, many independent observers find out the state of $S$ without Bohr-disturbing it or one another and one recovers Def. \ref{def:obj}. This gives the central result of the SBS approach:
\begin{theorem}[SBS]\label{ThmSBS}
\be
\text{Definition \ref{def:obj} $\Longleftrightarrow$ state (\ref{presbs}) with property (\ref{perp})},\nonumber
\ee
\end{theorem}
in the sense that (\ref{presbs},\ref{perp}) is the only state structure compatible with Def. (\ref{def:obj}). 
The above version of the SBS theorem together with its proof is a substantial generalization of the original result  from \cite{PRA} as nothing is assumed about the structure of the states ${\pmb \varrho_i}$ apart from (\ref{perp}). 
In the original SBS derivation \cite{PRA} an additional assumption of the, so called, strong independence was used: The only correlation between the parts of the environment is common information about the system. This forces  a fully product structure (cf. \cite{Feller}):
\be\label{strongindep}
{\pmb \varrho_i}= \varrho^1_i\otimes\cdots\otimes \varrho^k_i\otimes\cdots,
\ee
with $\varrho^k_i\varrho^k_{i'\ne i}=0$, which finally gives a state structure originally  known as SBS:
\be\label{SBS}
\varrho_{SBS}=\sum_i p_i \ket i \bra i \otimes\varrho^1_i\otimes\cdots\otimes \varrho^k_i\otimes\cdots, \quad \varrho^k_i\varrho^k_{i'\ne i}=0.
\ee
The only reason for adding the strong independence condition was admittedly to reproduce state structures found by the same authors in concrete models (e.g. in \cite{PRL}) and it is not needed for the objectivity per se as  Theorem \ref{ThmSBS} shows 
and as was also noted in \cite{Le}. Abandoning strong independence  leads to what has been called strong quantum Darwinism \cite{Le}. I will discuss it in the next Section. 

One important property has to be mentioned: SBS states (\ref{SBS}), or their more general form (\ref{presbs},\ref{perp}), imply quantum Darwinism condition (\ref{qD}) thanks to the orthogonality condition (\ref{perp}). Indeed, let us choose an arbitrary fraction $0<f\leq 1$ and consider a subset of $fN$ environments denoted $fE$. To calculate quantum mutual information of the reduced state $\varrho_{S:fE}$, first note that by (\ref{presbs}):
\ben
&&\varrho_{S:fE} = \sum_ip_i
\ket i \bra i \otimes \tr_{E \setminus fE}{\pmb \varrho_i}\\
&& \equiv \sum_ip_i
\ket i \bra i \otimes {\pmb \varrho_{fi}}
\een
Let us now calculate  all the necessary von Neumann entropies one by one, following the definition (\ref{qmi}). $\h(S) \equiv \h(\varrho_S)=\h(\sum_i p_i \ket i \bra i)=H(p_i)$ (the Shannon entropy of $p_i$). $\h(fE)\equiv \h(\tr_S \varrho_{S:fE}) =\h(\sum_i p_i {\pmb \varrho_{fi}})$. Since by (\ref{perp}) all ${\pmb \varrho_{fi}}$ have orthogonal supports for different $i$, we can easily write down the spectral decomposition of ${\pmb \varrho_{fi}}$, namely ${\pmb \varrho_{fi}}=\sum_\alpha \lambda_{i\alpha} \ket{{\pmb \psi_{i\alpha}}}\bra{{\pmb \psi_{i\alpha}}}$, where $\bra{{\pmb \psi_{i\alpha}}}{\pmb \psi_{j\beta}}\rangle=\delta_{ij}\delta_{\alpha\beta}$ and $\lambda_{i\alpha}\geq 0$, $\sum_\alpha \lambda_{i\alpha} = 1$ for every $i$. Thus one obtains:
\ben
&&\h(fE)=\h(\sum_i p_i \sum_\alpha \lambda_{i\alpha} \ket{{\pmb \psi_{i\alpha}}}\bra{{\pmb \psi_{i\alpha}}})\\
&& = -\sum_{i,\alpha} p_i \lambda_{i\alpha} \log(p_i \lambda_{i\alpha}) = H(p_i)+\sum_i p_i H(\lambda_{i\alpha})\\
&& \equiv \h(S) +\sum_i p_i \h({\pmb \varrho_{fi}}).
\een
Calculation of the remaining entropy $\h(\varrho_{S:fE})$ gives exactly the same result since the spectral decomposition of $\varrho_{S:fE}$ reads $\varrho_{S:fE} = \sum_i p_i \sum_\alpha \lambda_{i\alpha} \ket i \bra i \otimes \ket{{\pmb \psi_{i\alpha}}}\bra{{\pmb \psi_{i\alpha}}}$. Thus from (\ref{qmi}):
\ben
&&\I(S:fE) = \h(S) + \h(S) +\sum_i p_i \h({\pmb \varrho_{fi}})\nonumber\\
&& - \h(S) - \sum_i p_i \h({\pmb \varrho_{fi}}) = \h(S) \label{sbs qD}
\een 
and this holds for any fraction $f$. We thus reproduce the condition (\ref{qD}), identifying each environment $E_k$ with the critical fraction $f_0$ of (\ref{qD}).
(for more rigorous proof albeit in less general setting of (\ref{SBS}) but including error estimations see \cite{PRL, Mironowicz_PRL}).  Thus SBS, both in its general version (\ref{presbs}, \ref{perp}) and more particular (\ref{SBS}),  is a stronger condition than the quantum Darwinism  (\ref{qD}):
\be
SBS \Longrightarrow Quantum\ Darwinism.
\ee
But more importantly it is obtained using a very different philosophy: Through the formalization of Def. \ref{def:obj} rather than heuristic arguments. Quite surprisingly, this has lead to a unique  quantum state structure. In a sense, a philosophical notion of objectivity has been translated to quantum states. The question if the opposite implication holds, i.e. if (\ref{qD}) implies SBS turned out to be a difficult mathematical problem, solved in \cite{Le} and discussed in the next Section.

Interestingly, SBS states (\ref{SBS}) have been known in quantum Darwinism literature, at least in their simpler, pure version (see e.g. \cite{Zurek SciRep}). They appeared under the name of branching states (more precisely, branching states are quantum correlated states whose reductions are SBS). However, they were used only as examples of states satisfying quantum Darwinism condition (\ref{qD}) and their fundamental meaning as the only state structures compatible with the objectivity Definition \ref{def:obj} has not been recognized. It was only done in \cite{PRL, PRA}.

A search for SBS states in concrete applications is the hard part of the program as there are no universal tools yet allowing to obtain solutions for the dynamics of the central system plus a part of the environment. Master equation methods are of no use here due to the nature of the Born approximation, cutting the influence of the central system on the environment. The more appropriate limit is the recoiless limit, where the the recoil of the central system due to the environment is neglected. The search is then performed in two steps: i) first a part of the environment is neglected as unobserved and decoherence checked; ii) the next step is to check if the structure (\ref{SBS}) appears after  the decoherence and if the states $\varrho_i^k$ are orthogonal; the last test is performed using state fidelity as a convenient distinguishability measure \cite{Fuchs}. In many situations a single environment does not carry enough information and a grouping into so called macrofractions is preformed \cite{PRL} (see also \cite{Zurek_poczatki}).  The approach to SBS  is then governed, at least for pure dephasing Hamiltonians, by the approach theorem \cite{Mironowicz_PRL} bounding the trace distance of the actual state to the nearest SBS state by the decoherence factor and  pairwise state fidelities. 

By this method SBS states, or rather conditions for their formation, have been found in the most important models of open quantum system: Collisional decoherence \cite{PRL}, Quantum Brownian Motion in the recoiless limit \cite{Tuziemski qbm}, spin-spin systems \cite{Mironowicz_PRL, Mironowicz_PRA}, spin-boson systems \cite{Lewenstein} as well as in other models e.g. in a simple quantum electrodynamics model of a free charge \cite{QED} and in the recently proposed model of gravitational decoherence \cite{gravity}. Some interesting general results have also been obtained. Generic appearance of SBS for generalized von Neumann measurements was shown in \cite{measurements}, suggesting that objective measurement results are due to the SBS formation at the end of the measurement process. Also SBS  turned out to be more universal structures than bound to quantum theory and appear in a wide class of Generalized Probabilistic Theories (GPT) with a suitably defined objectivity notion\cite{GPT}.  In \cite{Kasia} an interesting fact was shown that in general entanglement is needed to produce SBS states. Finally, an intriguing dependence of SBS states on quantum reference frames was found in \cite{RF1, RF2}.

\section{Strong quantum Darwinism} \label{Sec:sqD}

That SBS states imply quantum Darwinism condition (\ref{qD}) is a simple calculation, however the opposite implication turned out to be a difficult mathematical problem. As was shown in Section \ref{Sec:cmi}, condition (\ref{qD}) it too weak to impose SBS structure. The question how to properly supplement it was solved by Le and Olaya-Castro in \cite{Le}, based on an intuition that the presence of discord is the obstacle (see also \cite{Pleasance,Le2018}).  This can inferred from Secs. \ref{Sec:nondisturb} and \ref{Sec:cmi}, but historically  the works \cite{Pleasance,Le2018,Le} where the first to explicitly name the problem. The original critique from \cite{PRA} was admittedly  vague and heuristic, which I tried to overcome in the previous  Sections.  Le and Olaya-Castro establish the following, highly non-trivial and technically difficult equivalence \cite{Le} (I present below a slightly improved formulation due to \cite{Feller}) :
\begin{theorem}[Strong quantum Darwinism]\label{thm:sqD}
Let $\h_E=\h_{E_1}\otimes\dots\otimes \h_{E_N}$. Then a state $\varrho_{S:E}$ is of SBS form (\ref{SBS}) if and only if the following conditions hold simultaneously:
\begin{eqnarray}
&& \I(S:E)=\chi(E|S)\label{sqD1}\\
&& \I_{acc}(S:E_k) = \h(S) \text{ for every } k \label{sqD2}\\
&& \I(E_1\cdots E_N|S)=0 \label{sindep2}
\end{eqnarray}
\end{theorem}

\noindent Above $\chi(E|S)$  is the Holevo quantity \cite{Nielsen Chuang} defined as:
\be
\chi(E|S)\equiv \max_{\Pi^S}\left[\h\left(\sum_i p_i\varrho_{E|i}\right)-\sum_i p_i\h(\varrho_{E|i})\right],
\ee
where the maximization is performed over rank-1 measurements $\Pi^S\equiv \ket i\bra i$ on $S$ and $\varrho_{E|i}\equiv \langle i|\varrho_{S:E}|i\rangle$ are the conditional post-measurement states of $E$. 
The quantity 
\be
\I_{acc}(S:E)\equiv \max_{M^S} \I[(M^S\otimes{\bf 1}) \varrho_{S:E}], 
\ee
is the so called accessible information, with the maximization performed over generalized measurements $M^S$ on $S$. Finally:
\be
\I(E_1\cdots E_N|S)\equiv \sum_{k=1}^N \h(E_k|S)-\h(E_1\cdots E_N|S),
\ee
is the multipartite mutual information, conditioned on $S$. The conditional von Neumann entropy is defined as: $\h(E|S)\equiv \h(SE)-\h(S)$.
Conditions (\ref{sqD1}, \ref{sqD2}) are called strong quantum Darwinism. From (\ref{d}, \ref{nondisturb_one}), (\ref{sqD1}) explicitly eliminates the one-sided discord $\delta(E|S)$ from the state $\varrho_{S:E}$. The second condition, (\ref{sqD2}) is the saturation of the accessible information - maximum information about $S$ can be recovered from any fragment of the environment. 
Finally, the last condition (\ref{sindep2}) is the correct mathematical expression of the strong independence condition \cite{Feller}, forcing the totally product structure (\ref{strongindep}).
As a result, after an elaborate mathematical derivation \cite{Le} (see also \cite{Feller}), one obtains that the only state structure compatible with all the above conditions is the SBS structure (\ref{SBS}).
As authors of \cite{Le} correctly point out, the strong independence (\ref{sindep2}) is not necessary per se for objectivity, i.e. it  does not follow from Definition \ref{def:obj} as the derivation in the previous Section shows. This was of course already known to the authors of \cite{PRA}, who pointed this fact out explicitly in the formulation of their main result (Theorem 1 of \cite{PRA}), but nevertheless added the condition to  be both: i) in agreement with at that time was discovered in models and ii) provide a sort of idealized structure where not only observers act independently, but the pieces of the environment they observe evolve independently too.  

Theorem \ref{thm:sqD} establishes the final missing link between the two approaches to objectivity: The information-theoretic of quantum Darwinism and the structural of SBS. It requires both a substantial modification of (\ref{qD}) and a relaxation of (\ref{SBS}) to (\ref{presbs}, \ref{perp}) if we drop strong independence. How to check for conditions (\ref{sqD1}, \ref{sqD2}) in practice is a separate, non-trivial problem due to e.g. the presence of the Holevo quantity. Some ideas can be found in \cite{Le2019}, but what is missing is some universal approach theorem, e.g. of the type \cite{Mironowicz_PRL}, which would allow for some error in (\ref{sqD1},\ref{sqD2}) and bound the resulting departure from the ideal SBS.

\section{Case study: Spin-spin model} \label{model}

As an example of how objectivization process happens, I will analyze the spin-spin model of decoherence, where a central spin interacts pair-wise with environmental spins. This is one of the  canonical model of quantum open systems \cite{open systems}, with a series of practical implications. I will analyze both the SBS formation and the condition (\ref{qD}), skipping the strong quantum Darwinism as the calculation of the accessible information and the Holevo quantity is in general a very complicated task, beyond the scope of the present work. Spin-spin model has been extensively studied from the objectivity point of view in \cite{Zurek spins,Mironowicz_PRA,Mironowicz_PRL, Kwiatkowski}. The Hamiltonian in the simplest case of the quantum measurement limit is just the interaction Hamiltonian:
\be\label{H1}
H=\frac{1}{2}\sigma_z\otimes\sum_{k=1}^N g_k\sigma_z^{(k)},
\ee
where $k$ enumerates the environment spins, $\sigma_i$ are Pauli matrices, with $\sigma_z^{(k)}$ denoting the matrix acting in the space of the $k-th$ spin and $g_k$ are coupling constants. 
The dynamics can be easily solved by diagonalizing the central spin observable $\sigma_z=\sum_{m=\pm 1}m\ket m\bra m$, so that:
\be\label{H1}
H=\sum_{m=\pm 1}\ket{m} \bra{m} \otimes \sum_{k=1}^N m g_k \sigma_z^{(k)}.
\ee
We then find that the evolution $U_{S:E}=e^{-itH}$ is of so called controlled-unitary type, where the state of the central system controls which unitary is applied to the environment: 
\be\label{USE}
U_{S:E}=\sum_{m=\pm 1}\ket{m} \bra{m} \otimes \bigotimes_{k=1}^N U_m^{(k)}(t),
\ee
where
\be\label{Um}
U_m^{(k)}(t)\equiv e^{-itm g_k \sigma_z^{(k)}}
\ee
We assume a totally uncorrelated initial state as we will be interested in the system-environment correlations buildup and do not want to inject any initial correlations:
\be\label{rhoSE0}
\varrho_{S:E}(0)=\sigma_{0S}\otimes \bigotimes_{k=1}^N \varrho_{0k}.
\ee
After evolving for time $t$ and discarding a part of the environment assumed to be unobserved, we are left with the so called partially traced state:
\begin{eqnarray}
&&\varrho_{S:fE}(t)=\tr_{(1-f)E} \left[ U_{S:E}\varrho_{S:E}(0)U_{S:E}^\dagger\right]\\
&&=\sum_{m=\pm 1} \alpha_{m} \ket{m} \bra{m}\otimes \bigotimes_{k\in fE} \varrho^{(k)}_m(t)\label{mama1}\\
&&+ \sum_m\sum_{m'\ne m} \alpha_{mm'} \Gamma_{mm'}(t) \ket{m} \bra{m'} \otimes \bigotimes_{k\in E_{obs}} U_m^{(k)} \varrho_{0k} U_{m'}^{(k)\dagger},\label{mama2}
\end{eqnarray}
where 
\be\label{Gamma}
\Gamma_{mm'}(t)\equiv \prod_{k\in (1-f)E}\tr\left[\varrho_{0k}U_{m'}^{(k)\dagger}(t) U_m^{(k)}(t)\right]
\ee
is the decoherence factor due to the unobserved environment $(1-f)E$, $\alpha_{mm'}\equiv \langle m| \sigma_{0S} |m'\rangle$, $\alpha_{m}\equiv\alpha_{mm}$ are the initial probabilities, and
\be\label{rhom}
\varrho_m^{(k)}(t)\equiv U_m^{(k)} \varrho_{0k} U_{m}^{(k)\dagger}.
\ee
The partially traced state $\varrho_{S:fE}(t)$  will be the main object of the study. 

Let  us begin the analysis with the SBS states: Under which conditions  $\varrho_{S:fE}(t)$ comes close to the nearest SBS state, so that the spin value $m$ of the central spin becomes SBS objective? Intuitively, from (\ref{mama1},\ref{mama2}) it should happen when: i) the coherent part (\ref{mama2}) disappears and ii) the states $\varrho_m^{(k)}(t)$ become distinguishable for $m\ne m'$.  In \cite{Mironowicz_PRL} this intuition was formalized and it was proven using standard state-discrimination techniques that for measurement Hamiltonians of the type (\ref{H1}), the approach, in the trace norm, to the nearest SBS state is controlled by the expression:
\ben
			&& \frac{1}{2} \min ||\varrho_{S:E_{obs}}(t) - \varrho_{SBS}||_{tr} \leq  \sum_{m\ne m'} |\alpha_{mm'}| |\Gamma_{mm'}(t)|\nonumber\\
			&&+\sum_{m\ne m'} \sqrt{\alpha_m\alpha_{m'}}\sum_{k\in (1-f)E} F \left( \varrho_{m}^{(k)}(t), \varrho_{m'}^{(k)}(t) \right),\label{approach}\\
\een
where
\be\label{fid}
F(\varrho, \sigma)\equiv\tr\sqrt{\sqrt\varrho\sigma\sqrt\varrho},
\ee
is the state fidelity, appearing here as a measure of distinguishability \cite{Fuchs}. Indeed, one can easily show that $\varrho, \sigma$ have orthogonal supports, and hence are one-shot perfect distinguishable, if and only if $F(\sigma, \varrho)=0$. Thus, in practical situations the approach to SBS is controlled by two functions: The usual decoherence factor (\ref{Gamma}) and 
the state fidelity $F \left( \varrho_{m}^{(k)}(t), \varrho_{m'}^{(k)}(t) \right)$. Both can be easily calculated for spin-$1/2$ systems as all the matrices can be computed explicitly and there is only one decoherence and fidelity factor $\Gamma_{+-}, F_{+-}$. (for higher spins see \cite{Kicinski}).For simplicity, let us assume that all the initial environmental states are the same $\varrho_{0k} \equiv \varrho_{0}$. Parametrizing $\varrho_{0}$ using the usual Euler angles of $SU(2)$,
\be\label{rho0}
\varrho_0=R(\alpha,\beta,\gamma)\text{diag}[\lambda, 1-\lambda]R(\alpha,\beta,\gamma)^\dagger,
\ee
we find easily that:
\be\label{G+-}
\Gamma_{+-}(t) = \prod_{k\in (1-f)E} \left[\cos(g_k t)+i(2\lambda-1)\cos\beta\sin(g_kt)\right].
\ee
Computation of the fidelity starts by noting that:
\ben
&&F \left( \varrho_{m}^{(k)}(t), \varrho_{m'}^{(k)}(t) \right)=\nonumber\\
&&\tr\sqrt{\sqrt{\varrho_0}U_{m}^{(k)\dagger}(t) U_{m'}^{(k)}(t)\varrho_0U_{m'}^{(k)\dagger}(t) U_m^{(k)}(t)\sqrt{\varrho_0}}.
\een 
To calculate the eigenvalues of the matrix $M$ inside the square root, the best is to use the fact that the eigenvalues satisfy $\lambda_++\lambda_-=\tr M$, $\lambda_+\lambda_-=\det M=1$. This gives:
\be
F_{+-}^{(k)}(t)=\sqrt{1-(2\lambda-1)^2\sin^2\beta\sin^2(g_kt)}.
\ee
It is clear form the above formula that the fidelity will not vanish as it is a periodic function of time. This motivates the introduction of a form of coarse-graining \cite{PRL}: The observed part of the environment is divided into fractions, called macrofractions. As there are no direct interactions between the environmental spins, we see from (\ref{mama1}) that the state of each macrofraction is a product one: $\varrho_m^{mac}(t)\equiv \bigotimes_{k\in mac}\varrho_m^{(k)}(t)$. A very useful property of the fidelity function is that it factorizes w.r.t. the tensor product so that the macrofraction fidelity becomes:
\be\label{Fmac}
F_{+-}^{mac}(t)=\prod_{k \in mac}\sqrt{1-(2\lambda-1)^2\sin^2\beta\sin^2(g_kt)}.
\ee
If moreover $g_k$ are randomized, which is the standard trick in spin environments to induce decoherence \cite{open systems}, the product above contains random phases which will average out to a typically small value for a big enough macrofraction. The same of course applies to the decoherence factor (\ref{G+-}). 
Note that the ranges of products in (\ref{G+-}) and (\ref{Fmac}) pertain to different parts of the environment: It is the traced out (unobserved) part in the former and a fraction of the observed part in the latter. Inserting (\ref{G+-}), (\ref{Fmac}) into (\ref{approach}) gives the final estimate of the SBS proximity. Its further analysis must be in general done numerically (see however \cite{Mironowicz_PRL,Mironowicz_PRA}) and will be presented later, after I analyze the quantum Darwinism condition. 

The analysis of (\ref{qD}) in the spin-spin model was performed in \cite{Zurek spins}. From the definition of quantum mutual information, we obtain:
\be\label{IfE}
\I(\varrho_{S:fE})=\h(\varrho_S)+\h(\varrho_{fE})-\h(\varrho_{S:fE}),
\ee
where as before $\h$ denotes von Neumann entropy. Let us start with the last term. Using the specific form (\ref{USE}) and the fully product structure of the initial state (\ref{rhoSE0}), we may manipulate the form of $\varrho_{S:fE}$ in the following way:
\ben
&&\varrho_{S:fE}=\tr_{(1-f)E}\left[U_{S:E}\varrho_{S:E}(0)U_{S:E}^\dagger\right]=\nonumber\\
&&U_{S:fE}\Big[\tr_{(1-f)E}\left(U_{S:(1-f)E}\varrho_{S:(1-f)E}(0)U_{S:(1-f)E}^\dagger\right)\\
&&\otimes\varrho_{fE}(0)\Big]U_{S:fE}^\dagger,
\een
where the obvious notation was used, e.g. $U_{S:fE}\equiv \sum_{m=\pm 1}\ket{m} \bra{m} \otimes \bigotimes_{k\in fE} U_m^{(k)}(t)$, $\varrho_{fE}(0)\equiv  \bigotimes_{k \in fE} \varrho_{0k}$ and similarly for $(1-f)E$. Using the facts that  von Neumann entropy does not change under the unitary rotation and $\h(\varrho\otimes \sigma)=\h(\varrho)+\h(\sigma)$, we obtain:
\be
\h(\varrho_{S:fE})=\h(\varrho_{fE}(0))+\h(\tilde\varrho_S),
\ee
where $\tilde\varrho_S\equiv\tr_{(1-f)E}\left(U_{S:(1-f)E}\varrho_{S:(1-f)E}(0)U_{S:(1-f)E}^\dagger\right)$. It differs from $\varrho_S = \tr_E \varrho_{S:E}$ only in the smaller size of the traced out environment. The entropy of both $\varrho_S$ and $\tilde\varrho_S$ can be easily calulated from (\ref{mama1}, \ref{mama2}) and reads:
\ben
&&\h(\varrho_S)=h(\lambda^E),\\
&&\lambda^E\equiv\nonumber\\
&& \frac{1}{2}\left(\alpha_++\alpha_-+\sqrt{(\alpha_+-\alpha_-)^2+4|\alpha_{+-} \Gamma_{+-}^E|^2}\right),
\een
where $h(x)\equiv -x\log x - (1-x)\log (1-x)$ is the binary entropy, $\alpha$'s denote the initial state parameters (cf. \ref{mama1}, \ref{mama2}), and $\Gamma_{+-}^E$ is the decoherence factor due to the whole of the environment, i.e. the product in (\ref{Gamma}) runs over the whole of $E$. The expression for $\h(\tilde\varrho_S)$ is similar with the decoherence factor reduced to $(1-f)E$, i.e. given by (\ref{Gamma}). The term $\h(\varrho_{fE}(0))$ is easy calculated as well due to the product structure of $\varrho_{fE}(0)$:
\be
\h(\varrho_{fE}(0))=fN \cdot h(\lambda),
\ee
where $\lambda$ is the eigenvalue of the initial state (\ref{rho0}). The only problematic term left is $\h(\varrho_{fE})$ and not much can be done here (although see \cite{Zurek spins, Mironowicz_PRA, Mironowicz_PRL})) but a numerical analysis to which I now turn.

The assumptions for the numerics are as follows. For definiteness, the environment is divided such that first $fN$ spins are the observed fraction $fE$ and the rest, $(1-f)E$, is the traced over. This represents a toy-model for SBS with only one macrofraction. The system initial state is pure:
\be
\ket{\psi_{0S}}=\frac{\ket{-1}+\ket{+1}}{\sqrt 2},
\ee
so that $\alpha_+=\alpha_-=\alpha_{+-}=1/2$. The environments are all initiated in the same state with $\lambda=0.1$, $\beta=5/8\pi$, cf. (\ref{rho0}). The coupling constants are drawn from a uniform distribution over $[0,1]$.  The calculation of $\h(\varrho_{fE})$ is resource-consuming and was done only for a rather low number of spins $N=14$, given the available resources. Much better simulations are available in the original works \cite{Zurek spins}. 
\begin{figure}[htbp]
\includegraphics[width=\columnwidth]{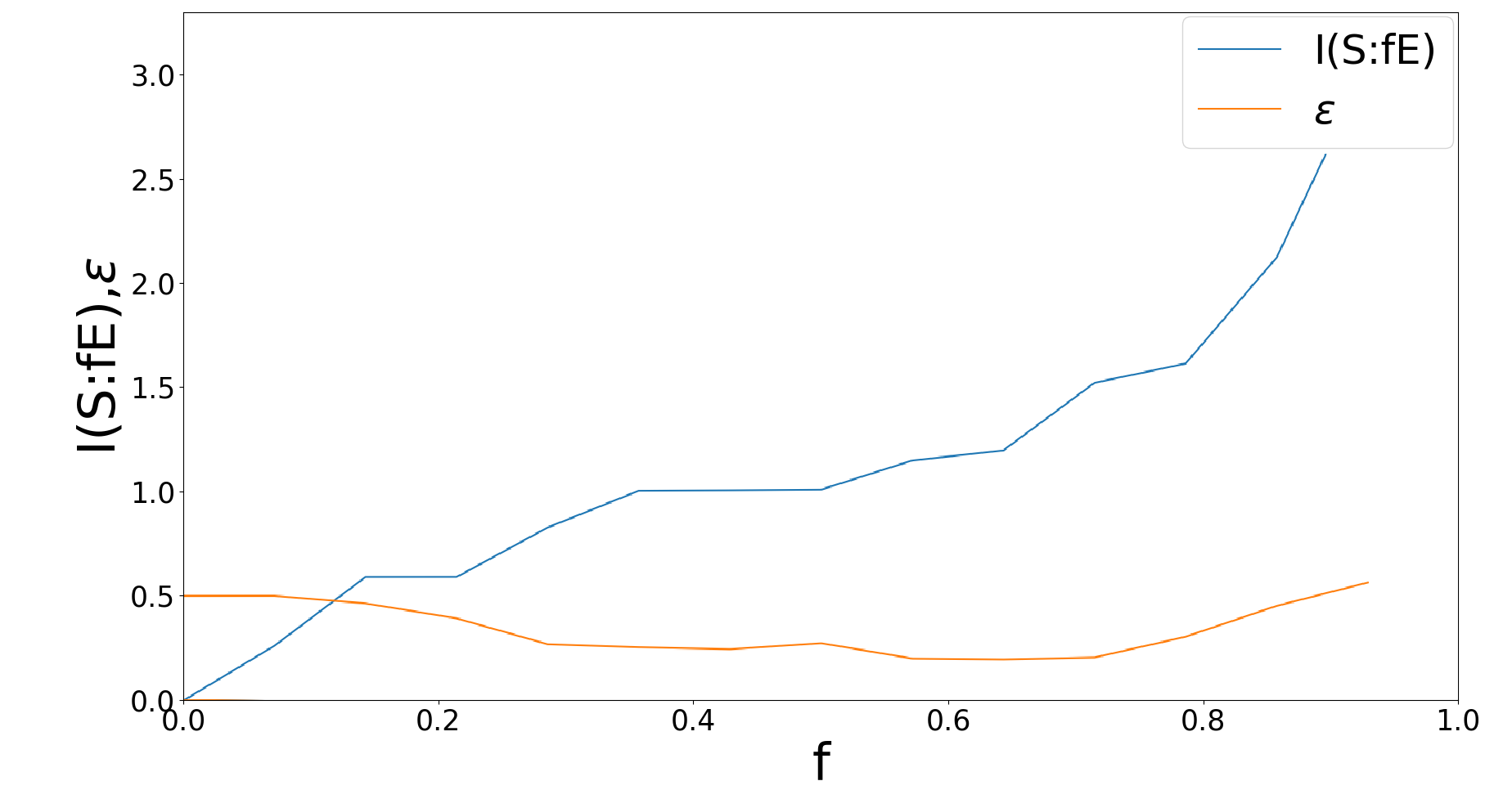}
\caption{Plots of $\I(\varrho_{S:fE})$ [blue curve] and the upper bound of the distance to the nearests SBS state (cf. (\ref{approach})), denoted by $\epsilon$ [orange curve] as a function of the observed fraction $f$ for a sample realization of the coupling constants $g_k\in[0,1]$ in the spin-spin model. The total number of spins is $N=14$, the time is set to $t=100$. The rest of the parameters are described in the text.}
\label{qD_SBS}
\end{figure} 
\begin{figure}[htbp]
\includegraphics[width=\columnwidth]{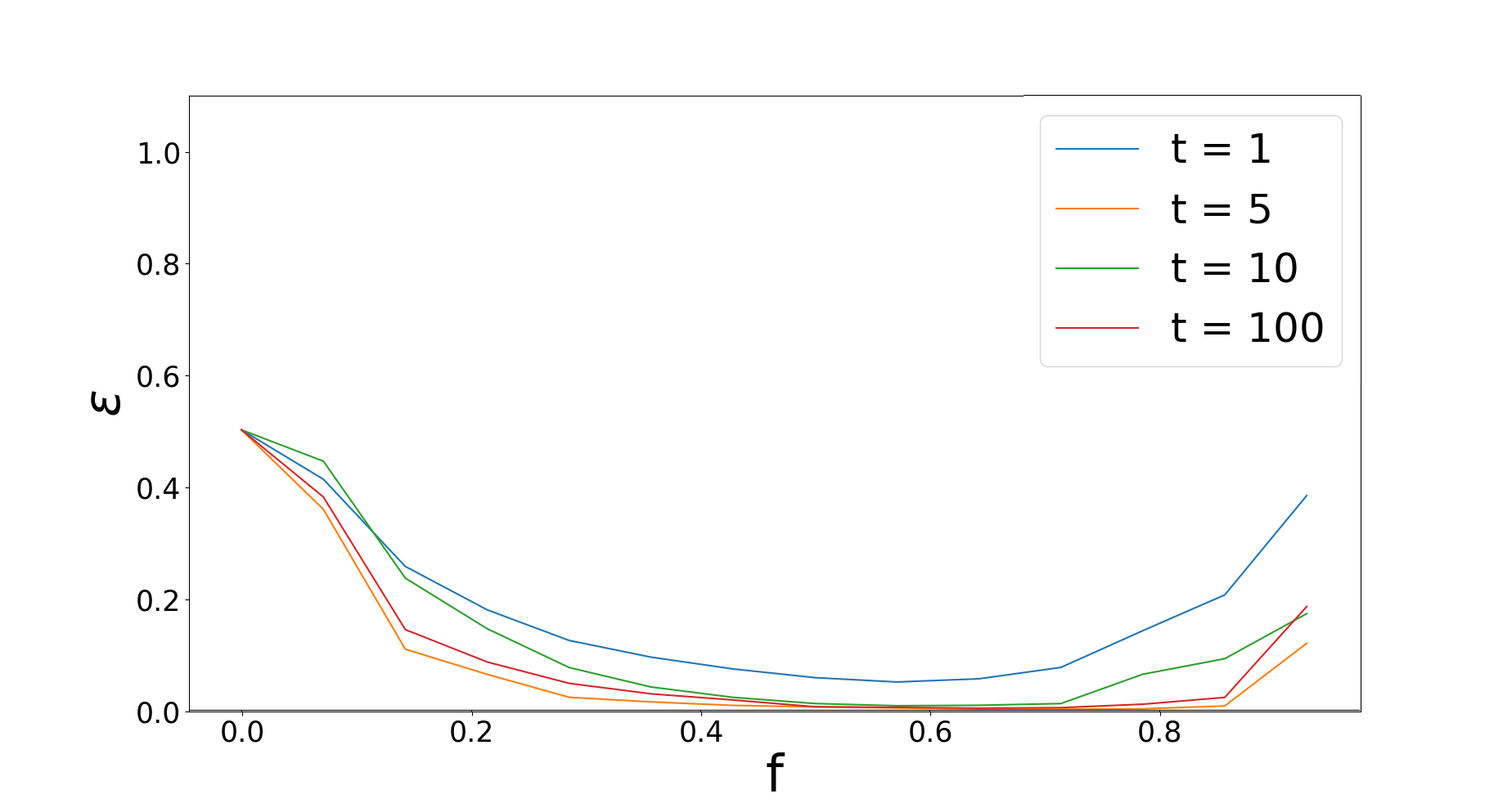}
\caption{Plots of  the upper bound of the distance to the nearests SBS state (cf. (\ref{approach})) for $N=50$ and different times. The rest of the parameters described in the text.}
\label{N50}
\end{figure} 
The results together with the SBS error function (\ref{approach}) are presented in Fig.~\ref{qD_SBS}. The plot of the distance to the nearest SBS, i.e. the right hand side of  (\ref{approach}) [orange curve] denoted here by $\epsilon$, shows an initial decrease with the increased $f$, which is due to the lowering of the fidelity function with the increasing dephasing due to the higher $fN$, cf. (\ref{Fmac}). From $f=0,3$ to approximately $f=0.7$ there is a slight recession, indicating the best, given the conditions, approach to a toy-SBS state with one macrofraction. It is then followed by an increase due to the increasing decoherence factor as less and less environments are traced over. The plot of $\I(\varrho_{S:fE})$ [blue curve], the partial information plot, is less obvious to interpret as the number of spins was too small to develop the proper classicality plateau. One can see a gentle onset of it after the initial increase past $f=0.3$ and before the rapid grow past $f=0.8$, when all possible quantum correlations start to enter $\I(\varrho_{S:fE})$. For higher $N$ the plots would have been more obvious \cite{Zurek spins} but it was beyond the available resources.

Thus in this example both methods show an onset of objectivity and approximately in the same region, but calculating the SBS  bound (\ref{approach}) has proven to be much more efficient, apart from the interpretation issues discussed earlier. In fact, as the upper bound in (\ref{approach}) can be calculated analytically, plotting it for larger $N$ does not present a problem. Sample results for $N=50$ are shown in Fig. ~\ref{N50}, clearly indicating an approach to a toy-SBS. 

\section{Comments on experimental tests of quantum Darwinism}\label{sec:exp}

Up to author's best knowledge, three experiments have been reported so far that aimed at directly testing for quantum Darwinism in the sense of condition (\ref{qD}) \cite{Ciampini, Chen, Jelezko}. All of them recently and with a very limited resources as the such investigations present a great experimental challenge. 
Of course the presented critique of the operational meaning of the condition (\ref{qD}) does not prevent it from being measured, i.e. by performing state tomography of the global state $\varrho_{S:E}$ and evaluating of $\I(S:fE)$ for the reconstructed density matrix. Let me briefly comment on these experiments, referring the reader to the original publications for more details.

In \cite{Ciampini} four-photon states were used to generate specific qubit graph states. One of the photons was chosen as the central system $S$, while three others played the role of the environment. By a proper state engineering, two graph states were generated: Star shaped, where the environmental photons were correlated only with the central photon and diamond shaped which had also intra-environment correlations. Although the resources where rather limited to only few qubits, they were enough to observe the condition (\ref{qD}) for the first case while not for the second. Let us closely look at the target state for the star shaped case. It is a modification of the well known cluster states \cite{cluster} and in the original notation of \cite{Ciampini} it reads:
\be
\ket{G_{N+1}} \equiv \prod_{j=2}^{N-2} \hat C(\theta_{j,j+1})\prod_{k=2}^N \hat C(\phi_{S,k}) \left (\ket +_S \otimes \bigotimes_{l=2}^N\ket +_l\right ),
\ee
where $\ket \pm =1/\sqrt 2(\ket 0 \pm \ket 1)$ and 
\be
\hat C(\phi_{j,k})\equiv \ket 0\bra 0_j\otimes {\pmb 1}_k+\ket 1\bra 1_j\otimes\left (\begin{array}{cc} 1 & 0\\ 0 & e^{i\phi_{j,k}}\end{array}\right )_k. 
\ee
Star shaped state corresponds to $\theta_{j,j+1}=0$, $\phi_{S,k}=\pi$ for all $j,k$. Let us pick any fraction of the environment, say first $n$ environmental qubits so that in our notation $fE=\{2,\dots,n+1\}$.
Then:
\begin{eqnarray}
&& \varrho_{S:fE}=\tr_{n+2,\dots,N} \ket{G_{N+1}}\bra{G_{N+1}}\\
&& =\frac{1}{2}\ket 0\bra 0\otimes \bigotimes_{k=n+2}^N\ket +\bra +_k + \frac{1}{2}\ket 1\bra 1\otimes \bigotimes_{k=n+2}^N\ket -\bra -_k\nonumber\\
&& + \frac{1}{2} \left(\underbrace{\langle -|+\rangle ^n}_0\ket 0\bra 1\otimes \bigotimes_{k=n+2}^N\ket +\bra -_k +\text{h.c.}\right)\\
&& = \frac{1}{2}\ket 0\bra 0\otimes \bigotimes_{k=n+2}^N\ket +\bra +_k + \frac{1}{2}\ket 1\bra 1\otimes \bigotimes_{k=n+2}^N\ket -\bra -_k,\nonumber
\end{eqnarray}
which is an SBS state (\ref{SBS}). Thus the reductions of the target experimental state were in fact SBS states and it comes as no surprise that the condition (\ref{qD}) was observed because of (\ref{sbs qD}). In the experiment state fidelity of at least $90\%$ was reported.

Similar situation took place in \cite{Chen}, where a six-photon quantum simulator was used to simulate the following global target state:
\begin{eqnarray}
&&\ket \Psi _{S:E}=\\
&&\alpha \ket 0_S \otimes\bigotimes_{i=1}^N \ket 0_i + \beta \ket 1_S \otimes\bigotimes_{i=1}^N \left[\cos\frac{\theta_i}{2}\ket 0_i+\sin\frac{\theta_i}{2}\ket 1_i\right]\nonumber.
\end{eqnarray}
The authors then simulated different angels $\theta_i$, controlling the overlap $\langle\theta_i|0\rangle$, and observed that for all $\theta_i=\pi$, the results followed the condition (\ref{qD}). Indeed, this is to be expected as reductions of $\ket \Psi_{S:E}$ read:
\begin{eqnarray}
&&\tr_{1,\dots,n}\ket\Psi\bra\Psi _{S:E}=\nonumber\\
&&|\alpha|^2 \ket 0\bra 0_S \otimes\bigotimes_{i=n+1}^N \ket 0\bra 0_i + |\beta|^2 \ket 1\bra 1_S \otimes\bigotimes_{i=n+1}^N \ket{\theta_i}\bra{\theta_i}\nonumber\\
&&+\alpha\beta^* \langle\theta_i|0\rangle^n  \ket 0\bra 1_S \otimes\bigotimes_{i=n+1}^N \ket 0\bra {\theta_i}+\text{h.c.}\label{chen_state}
\end{eqnarray}
where $\ket{\theta_i}\equiv \cos\frac{\theta_i}{2}\ket 0+\sin\frac{\theta_i}{2}\ket 1$. When $\langle\theta_i|0\rangle=0$ 
two things happen: i) decoherence - 
the coherent part in the last line of (\ref{chen_state}) vanishes; ii) orthogonalization - the residual environmental states $\ket 0$, $\ket {\theta_i}$ in the second line of (\ref{chen_state}) become orthogonal. Thus again a branching state, whose all reductions are SBS states (\ref{SBS}), was produced and the condition (\ref{qD}) follows from (\ref{sbs qD}). The authors also realize the power of the macrofraction method, introduced in in this context in \cite{PRL}, noting that even if $\langle\theta_i|0\rangle\ne0$, then by taking sufficiently large groups of qubits (macrofractions) their collective states become approximately orthogonal $\langle\theta^{mac}_i|0^{mac}\rangle=\langle\theta_i|0\rangle^n=\cos(\theta_i/2)^n\approx0$. The five 'environmental' photons  were actually meant to simulate  large and small  fractions by varying  $\theta_i$.

As a side note, for pure states of the environment, decoherence is equivalent to SBS as both are controlled by the same parameter - the decoherence factor. However for more realistic, noisy states of the environment, decoherence is by far not enough for SBS formation as orthogonalization must also take place (see e.g. \cite{measurements} for some generic SBS time scales evaluations). For example, a hot environment can be very efficient in decohering a system, while at the same time being so noisy that it carries practically no information about the system and thus no SBS can form (see e.g. \cite{Tuziemski qbm}).

Finally, in the third experiment \cite{Jelezko} nitrogen vacancy (NV) center in diamond was used as the central system with its nuclear spin surrounding as the environment.
Unlike in the previous experiments, the decohereing interaction here is natural,  given by the dynamics of the physical medium and not engineered.   
In a state-of-the-art experiment, four of the environmental spins were individually addressed and the Holevo quantity $\chi(fE|S)$ evaluated. The central result is the experimental demonstration that: 
\be\label{Jelezko}
\chi(fE|S)=\h(S)\  \text{\it independently of }  f,
\ee
from which quantum Darwinism is deduced, based on the operational interpretation of $\chi(fE|S)$ as the upper bound on a communication channel capacity. The fact that  only  the  part of (\ref{qD}) accessible via local measurements was considered (cf. (\ref{qmi decomp})), is probably the best exemplification of the problems with quantum mutual information discussed in Sections \ref{sec:qD} - \ref{Sec:cmi}. Indeed, the authors clearly state that they focus only on $\chi(fE|S)$, as the remaining part of the quantum mutual information - the discord, involves non-local correlations which do not help in establishing objectivity. This immediately rises the question if this non-local part is at all necessary? As the SBS approach shows, it is not. At least not in the sense of  Definition \ref{def:obj}. Note that condition (\ref{Jelezko}) is too weak to apply the Strong quantum Darwinism Theorem \ref{thm:sqD}, precisely because of the unknown discord, and hence one cannot deduce what was the underlying state structure. As a side remark, a theoretical study of SBS in NV centers has been undertaken in \cite{Kwiatkowski} in a hope to from an interface for  future SBS experiments.

\section{Conclusions}

I have presented here three approaches to what one can call 'the problem of objectivity', i.e. how to recover the objective character of macroscopic world from 
quantum theory. This is an important aspect of the famous quantum-to-classical transition that remained overlooked throughout the decades. More precisely, I have discussed three approaches to the proposed solution know as quantum Darwinism idea, which says that information becomes objective when it is 'fit enough' to not only survive the temporal evolution, but also to proliferate in some medium (called 'environment'). The first and the most popular approach is due to the original authors of the idea, W. H. Zurek and collaborators, and is based on the information-theoretical condition employing quantum mutual information (\ref{qD}). I have shown in detail that however intuitively appealing, it has a rather unclear operational interpretation in the quantum domain due to the generic presence of non-local correlations. This gave birth to a different approach, Spectrum Broadcast Structures, which to the contrary has a straightforward operational meaning. It builds directly from the fundamental definition of objectivity (Definition \ref{def:obj}) and encodes objectivity  in quantum state structure (\ref{SBS}), thus providing a reference to test against in models or experiments. Finally, the two approaches are joined by the third one, strong quantum Darwinism, which identifies what additional constraints must be imposed on the quantum mutual information to make (\ref{qD}) equivalent to (a somewhat generalized) Spectrum Broadcast Structure. It thus identifies what precisely information is  responsible for objectivity.

In author's view, the SBS program and its slight generalization to strong quantum Darwinism are the most promising lines of research as both approaches have a clear operational foundation. A working hypothesis behind the SBS program is that in realistic macroscopic situations, SBS states are notoriously generated leading to what we perceive as objective world. Some hints in that direction are presented in \cite{measurements}. There is still quite some work to be done, for example formulations of SBS theory for continuous variable systems (for initial ideas see \cite{gravity}) and SBS-objective motion are missing. So are experimental investigations of SBS in realistic situations.

\section*{Acknowledgements}

I would like to thank Mateusz Kici\'nski for a help with numerical simulations and preparing plots in Section\ref{model}. I acknowledge the support
by Polish National Science Center (NCN) through the grant
no. 2019/35/B/ST2/01896.

\bibliographystyle{apsrev}





\end{document}